\documentclass[aps,prd,twocolumn,preprintnumbers,superscriptaddress,floatfix]{revtex4}

\usepackage{multirow, graphicx,amssymb,url,mathrsfs,amsmath}
\usepackage{eucal,wrapfig,boxedminipage,setspace,subfigure}
\usepackage{amsxtra,amstext,latexsym,dsfont}

\def\IR{{\hbox{{\rm I}\kern-.2em\hbox{\rm R}}}}
\def\IB{{\hbox{{\rm I}\kern-.2em\hbox{\rm B}}}}
\def\IN{{\hbox{{\rm I}\kern-.2em\hbox{\rm N}}}}
\def\IC{\,\,{\hbox{{\rm I}\kern-.59em\hbox{\bf C}}}}
\def\IZ{{\hbox{{\rm Z}\kern-.4em\hbox{\rm Z}}}}
\def\IP{{\hbox{{\rm I}\kern-.2em\hbox{\rm P}}}}
\def\IH{{\hbox{{\rm I}\kern-.4em\hbox{\rm H}}}}
\def\ID{{\hbox{{\rm I}\kern-.2em\hbox{\rm D}}}}

\newcommand{\beq}{\begin{equation}}
\newcommand{\eeq}{\end{equation}}
\newcommand{\bea}{\begin{eqnarray}}
\newcommand{\eea}{\end{eqnarray}}

\begin{document}

\voffset 1cm

\newcommand\sect[1]{\emph{#1}---}

\title{Holographic Gauged NJL Model: the Conformal Window and Ideal Walking}

\author{Kazem Bitaghsir Fadafan}
\affiliation{ Faculty of Physics, Shahrood University of Technology,
P.O.Box 3619995161 Shahrood, Iran}

\author{Will Clemens}
\affiliation{ STAG Research Centre \&  Physics and Astronomy, University of
Southampton, Southampton, SO17 1BJ, UK}

\author{Nick Evans}
\affiliation{ STAG Research Centre \&  Physics and Astronomy, University of
Southampton, Southampton, SO17 1BJ, UK}

\begin{abstract}

\noindent We study the holographic Dynamic AdS/QCD description of a SU($N_c$) non-abelian gauge theory with $N_f$ fermions in the fundamental representation which also have
Nambu-Jona-Lasinio interactions included using Witten's multi-trace prescription. In particular here we study aspects of the dynamics in and near the conformal
window of the gauge theory as described by the two loop running of the gauge theory. If the number of flavours is such that the IR fixed point lies with the anomalous dimension, $\gamma$, of the quark bilinear above one then chiral symmetry breaking occurs. Here we display a spiral in the  mass - quark condensate plane describing a sequence of unstable excited states of the vacuum. An attractive NJL operator enhances the vacuum condensate but only an infinitely repulsive NJL interaction switches off the condensation completely. When $N_f$ changes so that the IR fixed point falls below one (the conformal window region)  there is a numerical discontinuity in the phase structure with condensation only occurring with a super critical NJL interaction.
In the conformal window, the running of $\gamma$ to a non-trivial IR fixed point is similar to walking dynamics, although chiral symmetry breaking is not triggered. In the ``Ideal Walking" scenario, chiral symmetry is broken in that IR conformal regime by the NJL interaction, but the change in $\gamma$ enhances the UV condensate. That enhancement of the condensate is shown in an analytic model with a sharp change in $\gamma$ and we show equivalent numerical results for the case of the two loop running. In the model the $\sigma$ becomes massless as the gauge theory running becomes near conformal and we show it is possible to realize a light higgs-like state in Ideal Walking models.

\end{abstract}

\maketitle

\newpage

Recently a holographic \cite{Maldacena:1997re} description of the gauged Nambu-Jona-Lasino (NJL) model \cite{Yamawaki:1996vr,Nambu:1961tp} was developed \cite{Clemens:2017udk}. The gauge theory action is
\begin{equation} {\cal L} = {1 \over 4 g_{YM}^2} F^{\mu \nu} F_{\mu \nu} + i \bar{q} D \hspace{-0.25cm} \slash  ~~ q  + {g^2 \over \Lambda^2} (\bar{q}_L q_R \bar{q}_R q_L + h.c.), \end{equation}
where $g_{YM}$ is the Yang-Mills coupling, $g$ the NJL coupling at the UV scale $\Lambda$, and where $N_f$ and $N_c$ can take any values.

Our analysis of the gauge theory is based on the Dynamic AdS/QCD model \cite{Alho:2013dka,Evans:2013vca,Evans:next}, which is motivated by top down holographic descriptions of chiral symmetry breaking
\cite{Karch:2002sh,Filev:2007gb}. The Dirac-Born-Infeld action, associated with probe flavour branes in these models, essentially describes a scalar field in AdS$_5$ space dual to the quark condensate.  The dynamics of the gauge theory enters through the background metric and fields but reduces to an effective running anomalous dimension of the quark mass/condensate, $\gamma$, which in AdS is a radially dependent mass squared for the scalar \cite{Alvares:2012kr,Jarvinen:2011qe}. The models share the insight that chiral symmetry breaking is triggered if $\gamma=1$ when the Breitenlohner Freedman (BF) bound is broken in AdS$_5$ \cite{Breitenlohner:1982jf}.
It is irresistible to take this very simple model of the quark dynamics and replace the running $\gamma$ with a sensible guess for another theory where a full holographic description of the glue background does not exist (note such a background might include backreaction of the quarks themselves). The Dynamic AdS/QCD model is such a  model with the perturbative running for an SU($N_c$) gauge theory with $N_f$ quarks inserted. If one simply takes the two-loop expressions for $\gamma$ for $N_c=3$ and $N_f=2$ naively extended to the scale where $\gamma=1$, a very sensible description of QCD is obtained (inspite of the very simplified choice of operators involved in the dynamics and the neglect of any stringy physics of the dual) as we shown in \cite{Clemens:2017udk}.

The NJL four-quark interaction may be included using Witten's double trace prescription \cite{Witten:2001ua} and it was shown how the basic NJL second order transition can be achieved holographically in non-supersymmetric duals in \cite{Evans:2016yas}. In \cite{Clemens:2017udk} we studied the gauged NJL model for the QCD case $N_c=3$ and $N_f=2$ - here the gauge dynamics runs $\gamma$ to an IR pole and breaks chiral symmetry itself. We explored the enhancement of the symmetry breaking by the NJL interaction. We extended the analysis in \cite{Clemens:2017luw} to look at phenomenological models of electroweak symmetry breaking such as technicolour, extended technicolour and top condensation.

This is the holographic description of the gauge theory we will again use here. Now though we will study theories with a non-trivial IR fixed point value for $\gamma$, which exist amongst the two loop running results for different choices of $N_c$ and $N_f$ \cite{Appelquist:1996dq}.
Of course, if the fixed point is non-perturbative then the two loop running can only be considered as a parametrization of the running rather than a precise computation. In this paper we will study the holographic description of these theories in more detail where there is considerable interesting structure.

For $N_f < 2.6 N_c$ the two loop running has an IR pole and these are the QCD-like theories we have explored in \cite{Clemens:2017udk}. As $N_f$ grows above $2.6 N_c$ though an IR fixed point develops with the fixed point value of $\gamma = \gamma_*$ falling from infinity. When the fixed point value of $\gamma_*$ lies above one there is still chiral symmetry breaking. Here there is interesting structure in the mass versus quark condensate plane -  a spiral structure indicating the presence of excited states of the vacuum. These structures have previously been observed in D3/D7 models of magnetic field induced chiral symmetry breaking \cite{Filev:2007gb}, the alternative dual of the conformal window of \cite{Jarvinen:2011qe,Jarvinen2}, in the condensed matter models of \cite{Iqbal:2011aj} and more recently in holographic superconductor models \cite{BitaghsirFadafan:2018iqr}. Analysis of the effective potential shows these states are higher energy excitations of the true vacuum and indeed the $\sigma$ meson is tachyonic in all but the lowest energy vacuum. They correspond to condensation of radially excited states of the $\sigma$. These excited states have also been observed to play a role in BKT transitions where the instability to chiral symmetry breaking occurs due to tuning the AdS scalar mass through the Breitenlohner Freeman bound, when the tachyons for each of these vacua become degenerate so the phase transition is not mean field \cite{Iqbal:2011aj, Jarvinen:2011qe}. Their appearance in so many models, including those with symmetry breaking but no BKT transition, suggests they are a robust prediction of holographic models of symmetry breaking. Here for us they play a small role in understanding the vacua of the theory with a repulsive NJL term. In the presence of an attractive NJL operator the condensation in the true vacuum is enhanced. We explore whether the excited states of the theory give rise to meta-stable vacua in the presence of the NJL term although we do not find such states. A repulsive NJL term in the true vacuum reduces the gap but only an infinitely repulsive term completely switches condensation off - we discuss this physics also at the level of the effective potential. A brief analysis of similar ideas in the alternative holographic model of gauge dynamics in \cite{Jarvinen:2011qe} can be found in \cite{Jarvinen2}.

As $N_f$ approaches $\sim 4 N_c$ from below the fixed point value falls very close to $\gamma_*=1$ and we enter the so called ``walking" gauge theory regime \cite{Holdom:1981rm}. As one reduces $N_f$ (which it is helpful to think of as a continuous parameter, as it is at large $N_c$) the scale at which $\gamma=1$ is crossed falls sharply relative to some UV scale where the coupling is fixed across comparator theories. Formally there is a BKT transition as $\gamma_*$ falls to precisely one and we see the spiral in the mass-condensate plane contract into the origin of the plot. Here at the low chiral symmetry breaking scale the quark condensate which has dimension close to two goes as $\langle \bar{q} q \rangle \sim f_\pi^2$, on dimensional grounds,  where we use the pion decay constant, $f_\pi$, to set the scale of the chiral symmetry breaking dynamics. As one runs to the UV, the dimension of the condensate transitions at some intermediate scale, given by approximately the one loop pole scale $\Lambda_1$, and the condensate transforms to the dimension 3 $\langle \bar{q} q \rangle \sim f_\pi^2 \Lambda_1$. By arbitrarily tuning $N_f$ one can arbitrarily separate the scales $f_\pi$ and $\Lambda_1$ so that in the UV $\langle \bar{q} q \rangle / f_\pi^3 \rightarrow \infty$. This behaviour was first studied because of a possible role in suppressing flavour changing neutral currents in technicolour dynamics - increasing the condensate moves extended technicolour dynamics off to a higher scale. It is though also an interesting behaviour to study in the space of non-abelian gauge theories on its own.

In these walking theories it has been argued that the effective potential may be very flat as a result of the near conformality of the gauge theory and thus the mass of the higgs-like $\sigma$ particle may be parametrically light relative to $f_\pi$. Again there is a motivation in describing the observed visible higgs but also an intrinsic field theory property of interest here. Both the growth of the quark condensate and the light $\sigma$ have been observed for the walking case in the holographic model we will study here \cite{Alho:2013dka}.
We will build on this study here with the properties of a slightly different scenario - ``ideal walking" \cite{Fukano:2010yv,Rantaharju:2017eej}.

A problem with walking is that $N_f$ actually takes integer values and so, without going to the extremes of theories with a very large number of colours, $N_c$, it is unlikely any real theory is sufficiently fine tuned to give exactly $\gamma_{*}=1$. Ideal walking theories have been proposed as an alternative set up with some of the same gains. Here one studies a gauge theory, at larger $N_f$ that lives in the so called conformal window \cite{Dietrich:2006cm} (ie $N_f \geq 4 N_c$ and less than $N_f = 11 N_c/2$ where the fixed point $\gamma_*$ falls to zero and asymptotic freedom is lost) and has a true IR conformal fixed point. It never becomes strong enough to trigger chiral symmetry breaking itself. Instead a strongly coupled NJL interaction at a scale greater than $\Lambda_1$ is used to trigger chiral symmetry breaking. Here the fine tuning to set $f_\pi$ below $\Lambda_1$ is provided by the NJL coupling rather than $N_f$. Now, if at the IR fixed point the anomalous dimension is $0< \gamma_{*} < 1$, then $\langle \bar{q} q \rangle \sim f_\pi^{3- \gamma_{*}}$. In the UV it becomes the dimension 3 object $\langle \bar{q} q \rangle \sim f_\pi^{3- \gamma_{*}} \Lambda_1^{\gamma_{*}}$. Note here that as one tunes $f_\pi \ll \Lambda_1$ again we have $\langle \bar{q} q \rangle / f_\pi^3 \rightarrow \infty$. This will be the case for any theory in the conformal window no matter the size of $\gamma_*$. This is almost certainly achievable with discrete $N_f$ at small $N_c$.

We begin study of this regime by determining the phase structure of the model at fixed $N_c$ in the $N_f$ against NJL coupling, $g^2$, plane. For $N_f < 4 N_c$ there is always chiral symmetry breaking driven by the gauge theory. Above this value of $N_f$, where a BKT transition occurs in the massless theory with changing $N_f$, our analysis suggests there is a numerical discontinuity in the plane with chiral symmetry breaking then only occurring for $g^2$ greater than a finite critical coupling, rather than a smooth transition with the critical $g^2$ value growing from zero. There may be an exponential dependence of the condensate against NJL coupling but we have not seen numerical evidence for that.

Next we explore the enhancement of the UV condensate (or equally the suppression of any UV mass relative to its IR value which is also expected). First we study an idealized running where  $\gamma$ transitions from zero to a fixed point value at a sharp
scale, $\Lambda_1$. We simply match the form of the solution on the two sides of the discontinuity analytically and show that the expected increase in the condensate (decrease in the mass) is realized. We then demonstrate the same phenomena occurring numerically with the two loop gauge theory running (where the intermediate scale is less clearly defined) - here it is easiest to follow the leading term in the holographic solution (the mass) but the behaviour is clearly reproduced.

Finally one might again wonder whether a light $\sigma$ particle could emerge in this ideal walking setting to provide a different possibility for electroweak physics and its light higgs. Here again in our model we can produce an analytic result,  showing that when the running of $\gamma$ is slow the mass of the $\sigma$ indeed falls towards zero. If one places the scale at which the NJL model causes symmetry breaking in the very conformal IR regime then its mass can be very small (below that needed to describe the higgs). We show that for $N_f=12$ choices of the IR symmetry breaking scale and the UV cut off scale exist for which $m_\sigma/f_\pi \simeq 0.5$ and hence might form the basis of an electroweak model.

We will first introduce Dynamic AdS/QCD, next Witten's prescription for the four fermion operators, and then present our computations for the model in each $N_f$ range. Since the two loop runnings are not non-perturbatively trustworthy we again stress that we take them as simply parameterizations of functions for $\gamma$ that run logarithmically from zero in the UV to non-trivial fixed points in the IR. The precise form is not trustable so we will just present results for $N_c=3$ as a function of (potentially fractional) $N_f$ since these include runnings from UV asymptotic freedom to all values of IR fixed point and are thus representative of all $N_c$.

\vspace{-0.5cm}

\section{Dynamic AdS/QCD}
\vspace{-0.5cm}

The Dynamic AdS/CFT model \cite{Alho:2013dka,Evans:2013vca} is a variant of the original AdS/QCD model \cite{Erlich:2005qh} that includes some key aspects of top down modeling such as in the D3/D7 system. The action is
\begin{equation} \begin{array}{lcl}
S & = & -\int d^4x~ d \rho\, {\rm{Tr}}\, \rho^3
\left[  {1 \over r^2} |D X|^2 \right.  \\ &&\\
&& \left. \hspace{1cm}  +  {\Delta m^2(r) \over \rho^2} |X|^2   + {1 \over 2 \kappa^2} (F_V^2 + F_A^2) \right],
\label{daq} \end{array}
\end{equation}
Here $X$ is a field dual to the quark condensate $\bar{q} q$, and the vector and axial vector fields describe the operators $\bar{q} \gamma^\mu q$ and $\bar{q} \gamma^\mu \gamma_5 q$. The theory lives in a geometry
\begin{equation} ds^2 = r^2 dx_{3+1}^2 + {1 \over r^2} d\rho^2, ~~~~~~~r^2 = \rho^2 + |X|^2 \end{equation}
$\Delta m^2$ is a renormalization group scale/radially dependent mass term.
$X$, the field that describes the quark mass and condensate, is included in the definition of $r$ in the metric. If we simply write $\Delta m^2(\rho)$ then if this term trips us through the BF bound in some range of small $\rho$ then the instability would exist however large $|X|=L$ were to grow. If we instead promote $\rho$ in this term to $r$ then at sufficiently large $L$ the instability is no longer present and a full solution for $L$ is possible. In addition when one looks at fluctuations about the background embedding those states only know about the non-trivial embedding through the factors of $L$ in the metric - these terms communicate the existence of a gap to the bound states.  We are being inconsistent about higher order terms in $X$ but the key point is to keep the $X^2$ term that triggers the BF bound violation and the brutality of our other assumptions should be judged by the success of the model.

\begin{figure}[]
\centering
\includegraphics[width=8cm]{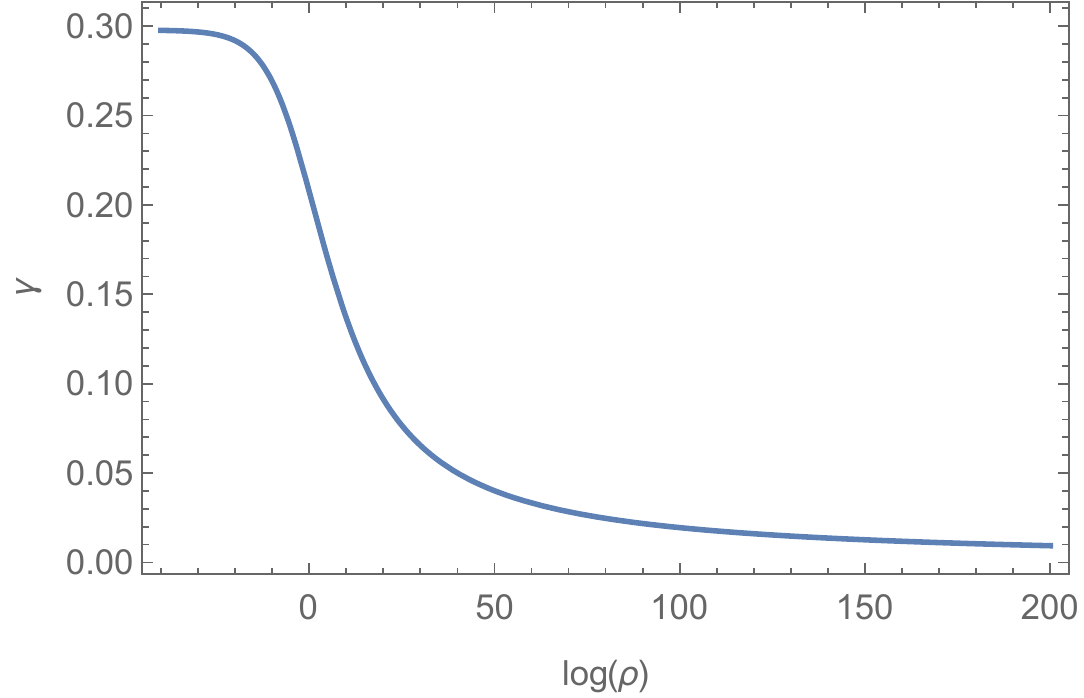}
\caption{An example plot of the running of $\gamma$, calculated from two loops, from the IR fixed point to the asymptotically free UV. $\gamma_{*}$ is the value in the IR. Here we have $N_c=3$ and $N_f=13$}
\label{gamvsmu}
\end{figure}

We will fix the form of $\Delta m^2$ using the two loop running of the gauge coupling in QCD with $N_f$ flavours transforming in the fundamental representation. This takes the form
\begin{equation}
\mu { d \alpha \over d \mu} = - b_0 \alpha^2 - b_1 \alpha^3,
\end{equation}
where
\begin{equation} b_0 = {1 \over 6 \pi} (11 N_c - 2N_F), \end{equation}
and
\begin{equation} b_1 = {1 \over 24 \pi^2} \left(34 N_c^2 - 10 N_c N_f - 3 {N_c^2 -1 \over N_c} N_F \right) .\end{equation}
Asymptotic freedom is present provided $N_f < 11/2 N_c$. There is an IR fixed point with value
\begin{equation} \alpha_* = -b_0/b_1\,, \end{equation}
which rises to infinity at $N_f \sim 2.6 N_c$.  Between this lower value and the higher value where asymptotic freedom is lost the theory has a smoothly decreasing IR fixed point. An example running is shown in Figure 1.

We will identify the RG scale $\mu$ with the AdS radial parameter $r$ in our model.
Working perturbatively from the AdS result $m^2 = \Delta(\Delta-4)$ \cite{Maldacena:1997re} we have
\begin{equation} \label{dmsq3} \Delta m^2 = - 2 \gamma = -{3 (N_c^2-1) \over 2 N_c \pi} \alpha\, .\end{equation}
The BF bound is violated when $N_f \simeq 4 N_c$ and theories at lower $N_f$ break chiral symmetry. The Conformal Window is the region between this transition and the loss of asymptotic freedom where the IR is an interacting fixed point.

The vacuum structure of the theory is found by setting all fields except $|X|=L$ to zero. The Euler-Lagrange equation for the determination of $L$, in the case of a constant $\Delta m^2$, is
\begin{equation} \label{embedeqn}
\partial_\rho[ \rho^3 \partial_\rho L]  - \rho \Delta m^2 L  = 0\,. \end{equation}
We can now ansatz the $r$ dependent $\Delta m^2$ above to describe the running of the dimension of $\bar{q}q$ (we do this at the level of the equation of motion).
To find numerical solutions, we need an IR boundary condition. In top down models $L'(0)=0$ is the condition for a regular solution.  Since we do not wish to describe IR physics below the quark mass (where the quark contribution to the running coupling will decouple) we use a very similar on-shell condition - we shoot from points $L(\rho=m_{IR}) = m_{IR}$ with $L'(m_{IR})=0$. In the UV the solution (neglecting $\Delta m^2$ which falls close to zero) takes the form
\begin{equation} L = m _{UV}+ {c_{UV} \over \rho^2} \label{UVform} \end{equation} where $m_{UV}$ in interpreted as the UV quark mass and $c_{UV}$ as the quark condensate. Formally to find the expectation value of the condensate one must substitute the solutions back into the action and differentiate with respect to $m_{UV}$ \cite{Evans:next}. There is a divergent piece but $c_{UV}$ controls the dynamically determined condensate. Note away from the very far UV locally the solution takes the form $L = m/\rho^{\gamma(\rho)} + c/ \rho^{2-\gamma(\rho)}$.

The spectrum of the theory is found by looking at linearized fluctuations of the fields about the vacuum where fields generically take the
form $f(\rho)e^{ip.x}, p^2=-M^2$. A Sturm-Louville equation results for $f(\rho)$ leading to a discrete spectrum. By substituting the wave functions back into the action and integrating over $\rho$ the decay constants can also be determined.
The normalizations of the fluctuations are determined by matching to the gauge theory expectations for the VV, AA and SS correlators in the UV of the theory. This full procedure is described in detail in \cite{Alho:2013dka}.
 In particular the scalar mode obeys
\begin{equation}\label{scalareom}
\partial_\rho (\rho^3S^{'})-\Delta m^2\rho S-\rho L_0S\frac{\partial \Delta m^2}{\partial L}+M^2\frac{\rho^3}{(L_{0}^2+\rho^2)^2}S=0
\end{equation}
with $S$ the scalar field, $L_{0}$ the base background solution and $M$ the mass of the scalar mode. We now look for solutions with the boundary conditions $S(\Lambda_{UV})=1/\rho^2$ and $S^{'}(m_{IR})=0$.  We find that these solutions only exist for a discrete set of values for $M$, corresponding to the spectrum of scalar mesons - the $\sigma, \sigma^*, \sigma^{**}..$.

The vector mass is calculated in a very similar way using the equation of motion
\begin{equation}
\label{vectoreom}
\partial_\rho (\rho^3V^{'})+M^2\frac{\rho^3}{(L_{0}^2+\rho^2)^2}V=0
\end{equation}
with $M$ the vector mass now and $V$ the vector field. The mass is then obtained by looking for solutions with the same boundary conditions as the scalar.

The pion decay constant $f_\pi$ that we will use to normalise our results is given by the formula
\begin{equation}
\label{fpi}
f_\pi^2=\frac{1}{\kappa^2}\int d\partial_\rho \left[\rho^3\partial_\rho K_A (q^2=0)\right] K_A (q^2=0)
\end{equation}
where $K_A (q^2=0)$ is an externally sourced, massless axial field on the branes.

With $N_c$ and $N_f$ fixed the free parameters in the theory are the overall scale $\Lambda_{1}$, the UV quark mass and the 5d coupling $\kappa$. For example, if one wishes to fit to $N_c=3, N_f=2$ QCD one can fix $\Lambda_{1}$ by scaling to give the correct $m_\rho$; the remaining parameter $\kappa$ can then be fitted to the data. In \cite{Clemens:2017udk} we found a best fit to QCD data with  $\kappa =8.7$ but it would be expected to vary with $N_f$. Here we will work with $\kappa=1$ as a reference value - $f_{\pi}$ grows as $\kappa$ grows but the phenomena we report on are qualitatively the same.

\section{NJL Interactions}
\vspace{-0.5cm}

Consider a free fermion with a four fermion interaction $g^2/\Lambda^2 \bar{q}_L q_R \bar{q}_R q_L$. In the standard NJL approximation there are two contributions to the effective potential \cite{Nambu:1961tp}. First there is the one loop Coleman Weinberg potential \cite{Coleman:1973jx} for the free quarks
\begin{equation} V_{\rm eff} = - \int^\Lambda_0 {d^4k \over (2 \pi)^4} Tr \log (k^2 +m^2) \label{ColemanW}\end{equation}
This falls with growing $m$ and is unbounded, although normally one treats $m$ as a fixed parameter so one would not seek to minimize this potential. When we add the four fermion term we allow $m$ to become dynamically determined but there is the second  term from the four fermion interaction evaluated on $m= (g^2/\Lambda^2) \langle \bar{q} q \rangle$
\begin{equation} \Delta V_{\rm eff} = {\Lambda^2 m^2 \over g^2} \label{NJLextra} \end{equation}
This makes the effective potential bounded and ensures a minimum. For small $g$ the extra term is large and the minimum is at $m=0$. When $g$ rises above $2 \pi$ the minimum lies away from $m=0$. The phase transition is second order.

In Witten's prescription for ``multi-trace'' operators \cite{Witten:2001ua} we add the equivalent of the extra potential term (\ref{NJLextra}) as a boundary term at the UV cut off $\Lambda$. For large $\Lambda$ where $L\simeq m$ the term we add is
\begin{equation} \Delta S _{UV}=   {L^2 \Lambda^2 \over g^2  } \label{extrabit} \,.  \end{equation}
The effective potential from the background model is computed by evaluating minus the action (\ref{daq}) evaluated on the vacuum solution as a function of the UV mass term. We extract the values of $m$ and $c$ in the UV by fitting to the form (\ref{UVform}) near the cut off.

We can also understand Witten's prescription in terms of a change to the UV boundary condition on the solution of the embedding equation. Varying the action gives
\begin{equation}  \delta S = 0 = -\int d \rho \left(\partial _\rho{\partial {\cal L }\over \partial L'}  - {\partial {\cal L }
\over \partial L} \right)  \delta L  + \left. {\partial {\cal L }\over \partial L'} \delta L \right|_{{UV, IR}} \,.  \end{equation}
Normally in the UV one would require the mass to be fixed and $\delta L=0$ to satisfy the boundary condition but now we allow $L$ to change and instead impose
\begin{equation} 0 = {\partial {\cal L} \over \partial L'}  + {2 L \Lambda_{UV}^2 \over g^2 }  \,,   \end{equation}
where we have included the variation of the surface term. For our action ${\partial {\cal L} \over \partial L'} = \rho^3 \partial_\rho L$. Assuming (\ref{UVform}) we find that we need
\begin{equation} m \simeq {g^2 \over \Lambda^2} c \label{cond} \end{equation}
This condition is simpler to apply to the solutions of the Euler Lagrange equation than constructing and minimizing the effective potential but equivalent.  We caveat this by noting that in the presence of a repulsive NJL interaction, which places a minus sign in the potential term in (\ref{extrabit}), the condition finds a local maximum of the effective potential - we discuss this below in more detail.

\begin{figure}[]
\centering
\includegraphics[width=8cm]{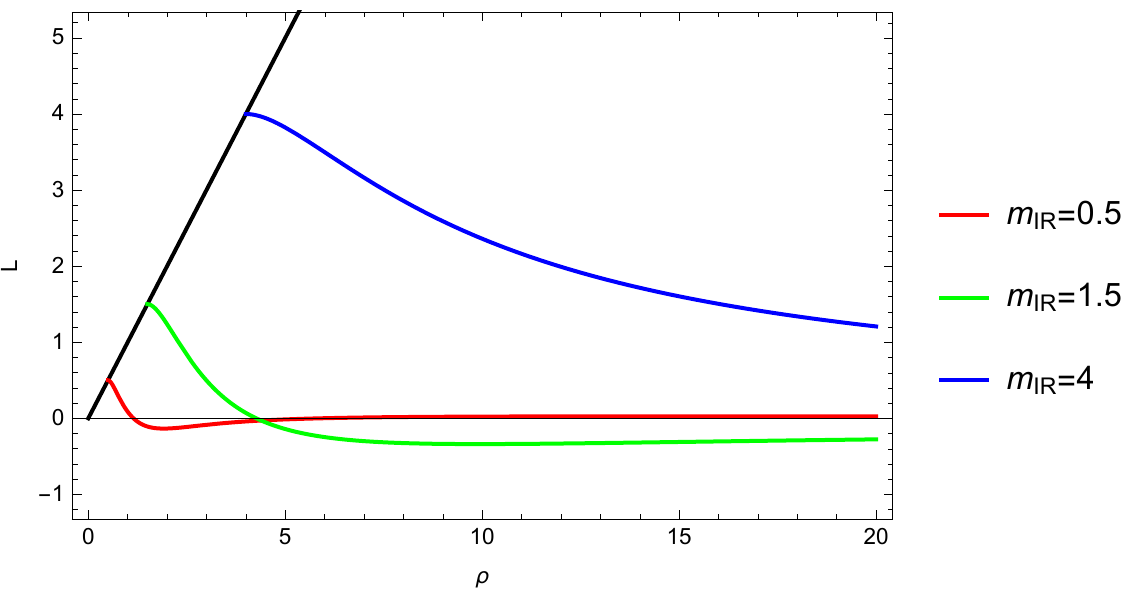}
\caption{The functions $L(\rho)$ with $m=0$ in the far UV for $N_f=9$. In the IR we cut off scales below where the quarks become on mass shell when $L(\rho=m_{IR})=m_{IR}$. Here the BF bound is violated at $r=11$.}
\label{spiral}
\end{figure}

\section{Chirally Broken Phase: $2.6 N_c < N_f < 4 N_c$}

In this range of $N_f$ the IR fixed point value of $\gamma$ is greater than one and the gauge theory generates chiral symmetry breaking on its own. To study this we seek solutions of (\ref{embedeqn})  with the IR boundary conditions described. All the dynamical scales are set in terms of the scale at which  $\gamma=1$ (which here we set to be at $r$=11))  and the UV value of $L$ which is the quark mass.  From the UV form of the solution (\ref{UVform}) we extract $m_{UV}$ and $c_{UV}$. In Figure 2 we plot the solutions for $N_f=9$ with $m_{UV}=0$ in the UV.

Note there are an infinite set of such solutions  - as the IR boundary value of $L$ shrinks solutions with more oscillations can be found. The solutions with more oscillations are excited states of the vacuum -  these are states where radially excited states of the $\sigma$ meson are condensed, which since they are all described by a single holographic field are mixed together. In Figure 3 we plot the position of the solutions
in the $m_{UV}-c_{UV}$ plane for two representative values of $N_f$ - to compare theories we have chosen to fix the UV scale in both where $\gamma=0.3$. As can be seen at $N_f=9$ (the top plot) there is a spiral structure. The spiral makes an infinite number of loops before ending at the origin. This structure has been previously observed in the D3/probe-D7 model with a magnetic field \cite{Filev:2007gb}, the alternative dual of the conformal window of \cite{Jarvinen:2011qe,Jarvinen2}, in the condensed matter models of \cite{Iqbal:2011aj} and more recently in holographic superconductors \cite{BitaghsirFadafan:2018iqr}
 so it appears very generic to holographic symmetry breaking descriptions.

\begin{figure}[]
\centering
\includegraphics[width=7cm]{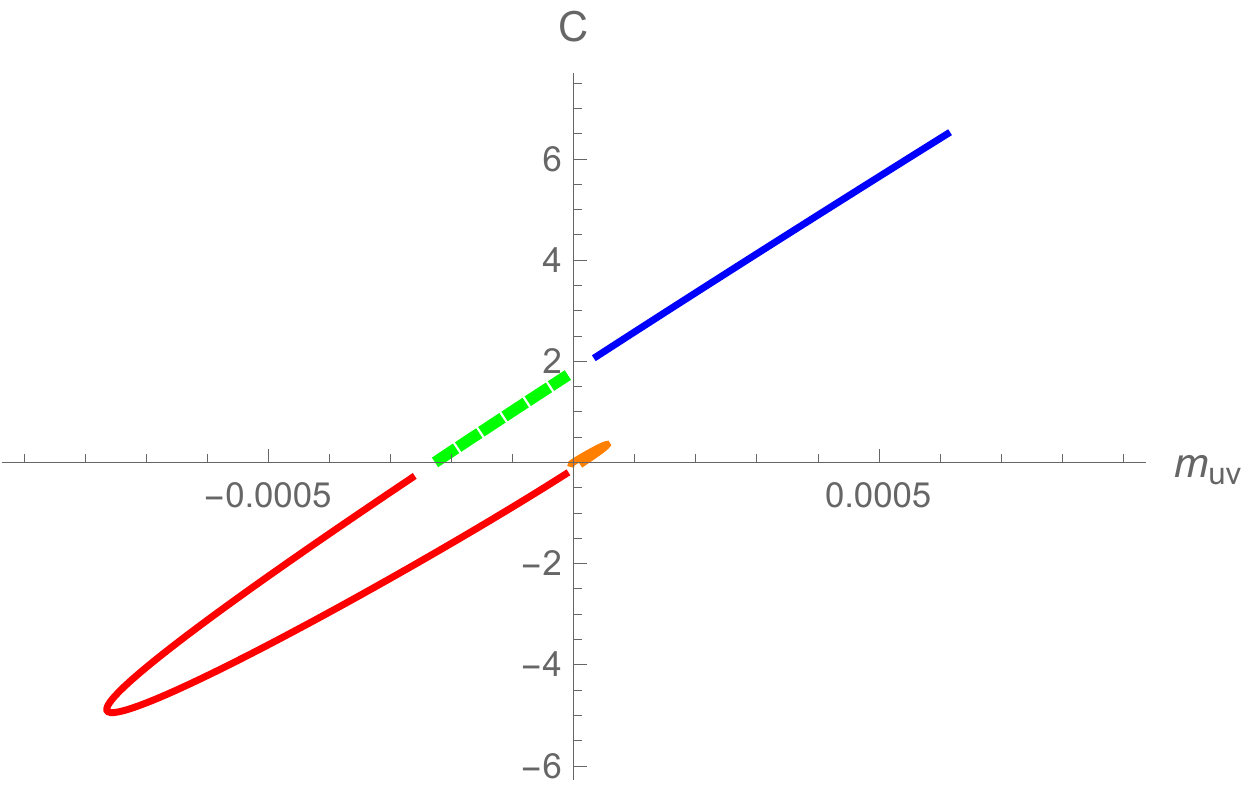}   \includegraphics[width=7cm]{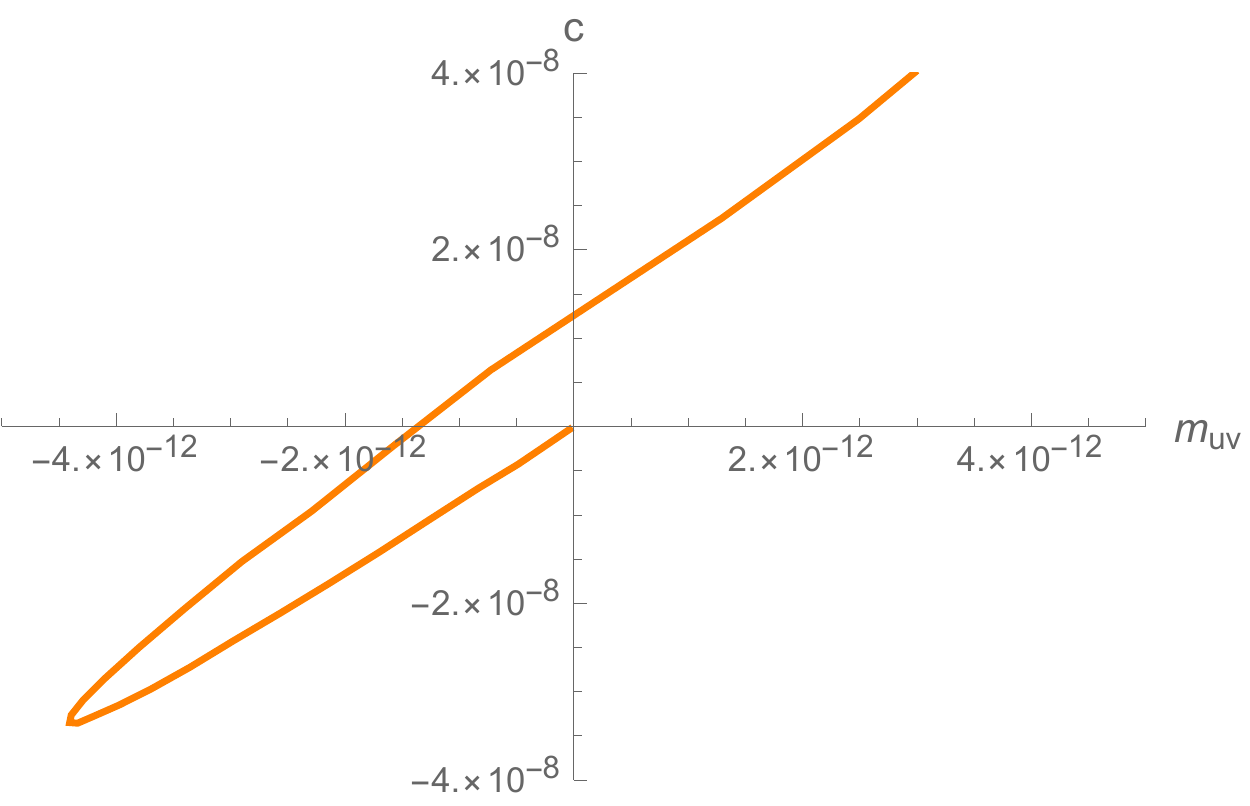}
\caption{The regular embeddings $L(\rho)$ plotted in the $m_{UV}-c_{UV}$ plane for (from top to bottom) $N_f=9,11$ showing the spiral structure and how the scale of chiral symmetry breaking shrinks as one approaches the BKT transition at $N_f \simeq 12$. Here both theories have $\gamma=0.3$ at the same UV scale.}
\label{spiral}
\end{figure}

The solutions of the Euler Lagrange equation represent turning points of the action and hence the effective potential ($V_{eff}=-S$ evaluated on the vacuum solutions). It is a simple matter to evaluate the vacuum energy on the solutions and show that they monotonically increase in energy with the number of axis crossings. The flat embedding $L=0$ has the highest energy. This means these turning points must be points of inflection of the effective potential since there are no interchanging maxima and minima. The interpretation is that the chirally symmetric phase $L=0$ is unstable to condensation of the $\sigma$ excitation but also all its radially excited states $\sigma^*, \sigma^{**}..$. We envisage a four dimensional low energy potential for these states of the form
\beq V_{eff} = -m_1^2 |\sigma|^2 - m_2^2 | \sigma_*|^2 + \lambda (|\sigma|^2 + |\sigma_*|^2)^2 + ... \eeq
This has minima of different depths on the $\sigma$ and $\sigma_*$ axes but the minima are smoothly connected in the full space with only one true local minimum where $\sigma$ alone has a vev  (see the top part in Figure 4 for a sketch of this form).

A simple way to test this hypothesis is to compute the $\sigma$ meson mass in each of the vacua. In the lower part of Figure 4 we plot the $\sigma$'s mass against the UV quark mass as we move along the spiral of Figure 3 and it is worth considering the physics here in detail. First note that at $m_{UV}=0$  only the $L$ profile that does not cross the $\rho$ axis (the top curve in Fig 2) has a stable $\sigma$. This is the true vacuum of the theory. There is another U(1)$_A$ rotated equivalent version of the vacua where the curve lies entirely below the axis (this is the opposite side of the wine bottle shaped potential one expects). Now as we switch on a positive $m_{UV}$ as expected the $\sigma$ mass rises.  When we add in a negative mass there is a period when the $\sigma$ mass remains positive - here we are describing the fate of the U(1)$_A$ rotated vacua in the presence of a mass. The vacuum manifold is tilted and this state is the unstable side of the potential (the pion is tachyonic). For sufficiently large negative mass the wine-bottle shaped form of the effective potential is lost and the non-true vacuum side of that potential becomes unstable for the $\sigma$ also. Beyond this point though the solution remains as a turning point of the potential - to understand its evolution we can follow it back to the $m_{UV}=0$ point where it is the second curve down in Fig 2. This is the vacuum where the $\sigma^*$ has condensed (although it is unstable to a roll to the true vacuum since the $\sigma$ is tachyonic). Tracking back along negative $m_{UV}$ tells us we are seeing this state in the presence of the negative mass. Continuing on one moves smoothly to the even less stable vacua with condensation of higher excitations of the $\sigma$.

\begin{figure}[]
\centering
\includegraphics[width=8cm]{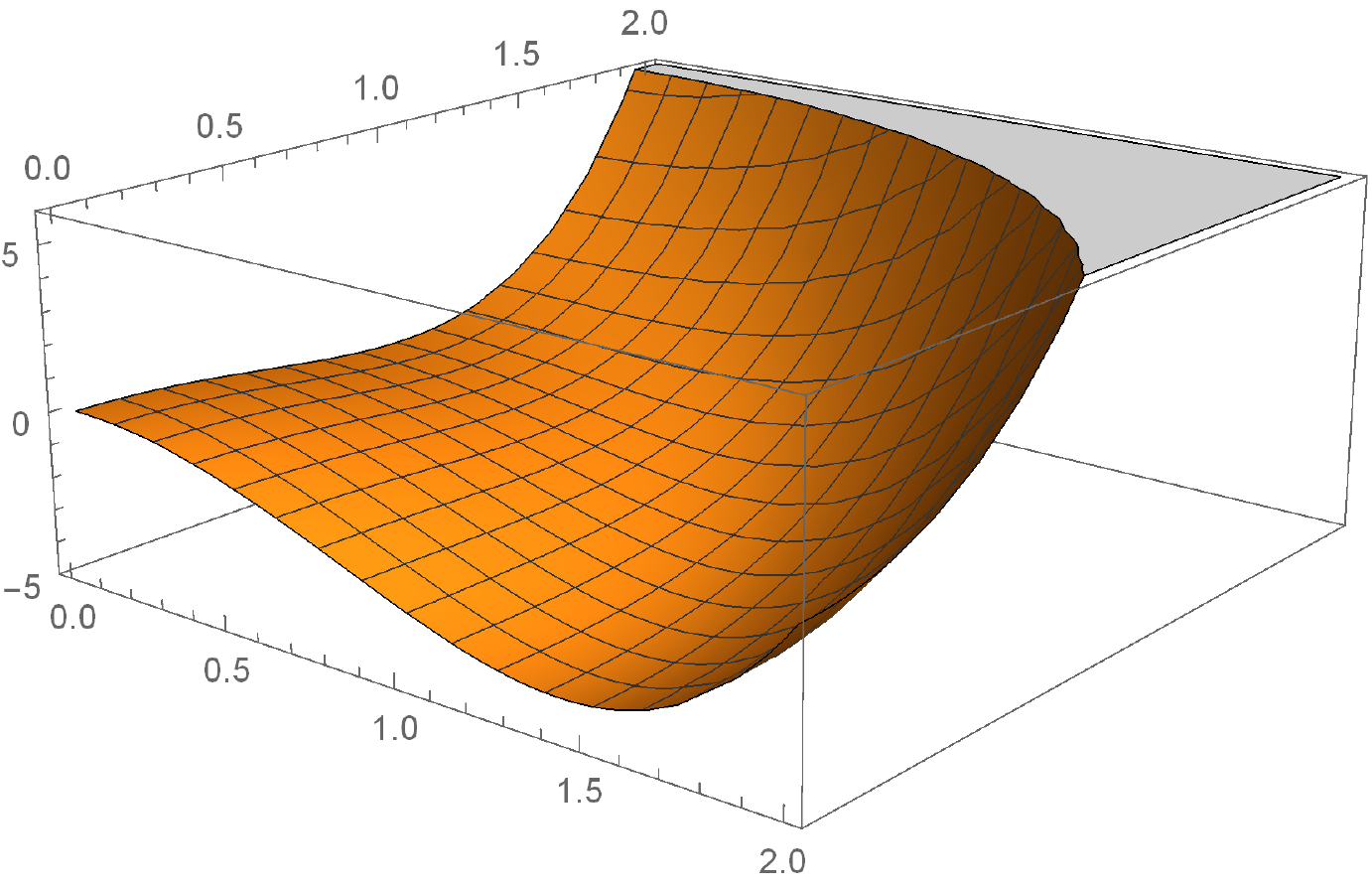}  \includegraphics[width=8cm]{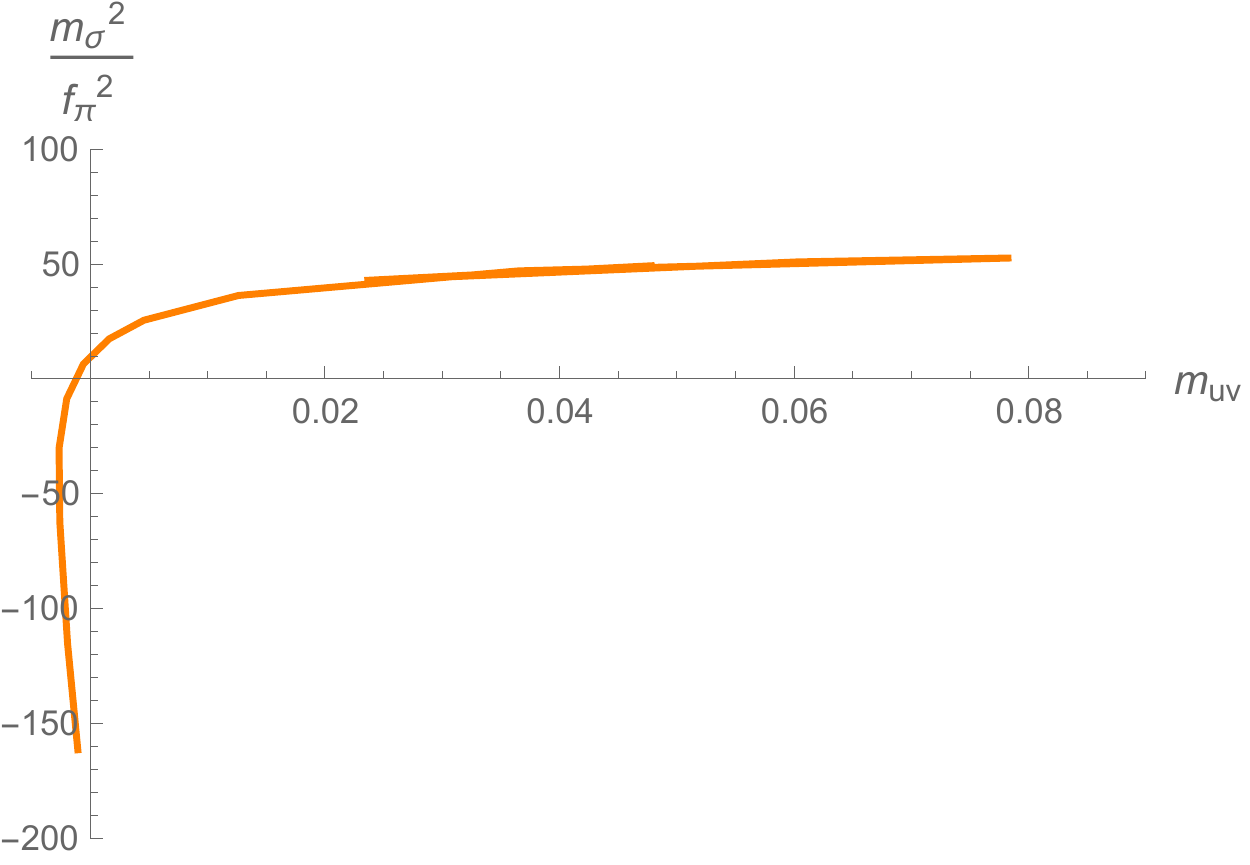}
\caption{Above - a sketch of the low energy potential against the $\sigma$ and $\sigma_*$ fields showing mimina on each axis but only a single true minima on the $\sigma$ axis. Below - the $\sigma$'s mass against the UV quark mass as we move along the spiral of Figure 3 with $N_f=9$ showing the instability of the excited states at $m_{UV}=0$.}
\label{metastable}
\end{figure}

These spirals have also been previously seen in the alternative model of the $N_f$ dependence of gauge theories in \cite{Jarvinen:2011qe,Jarvinen2}  and in the condensed matter models of \cite{Iqbal:2011aj}. There the authors stressed the role of these extra vacua as one approaches a BKT phase transition. If the AdS scalar mass can be tuned to the BF bound (as here by tuning $N_f$) then a non-mean field transition occurs because a set of Efimov states emerge - these correspond to an infinite number of tachyonic states whose masses pile up at zero and play a role in the transition. Here as $N_f$ approaches the chiral transition it is these tachyons, for rolls to each of the vacua for the excited states of the $\sigma$, that are these Efimov states. Here, the existence of these states will play a role in understanding the response of the theory to NJL interactions. A brief similar analysis in the alternative model of the conformal window of \cite{Jarvinen:2011qe} can be found in \cite{Jarvinen2}.

\begin{figure}[]
\centering
\includegraphics[width=8cm]{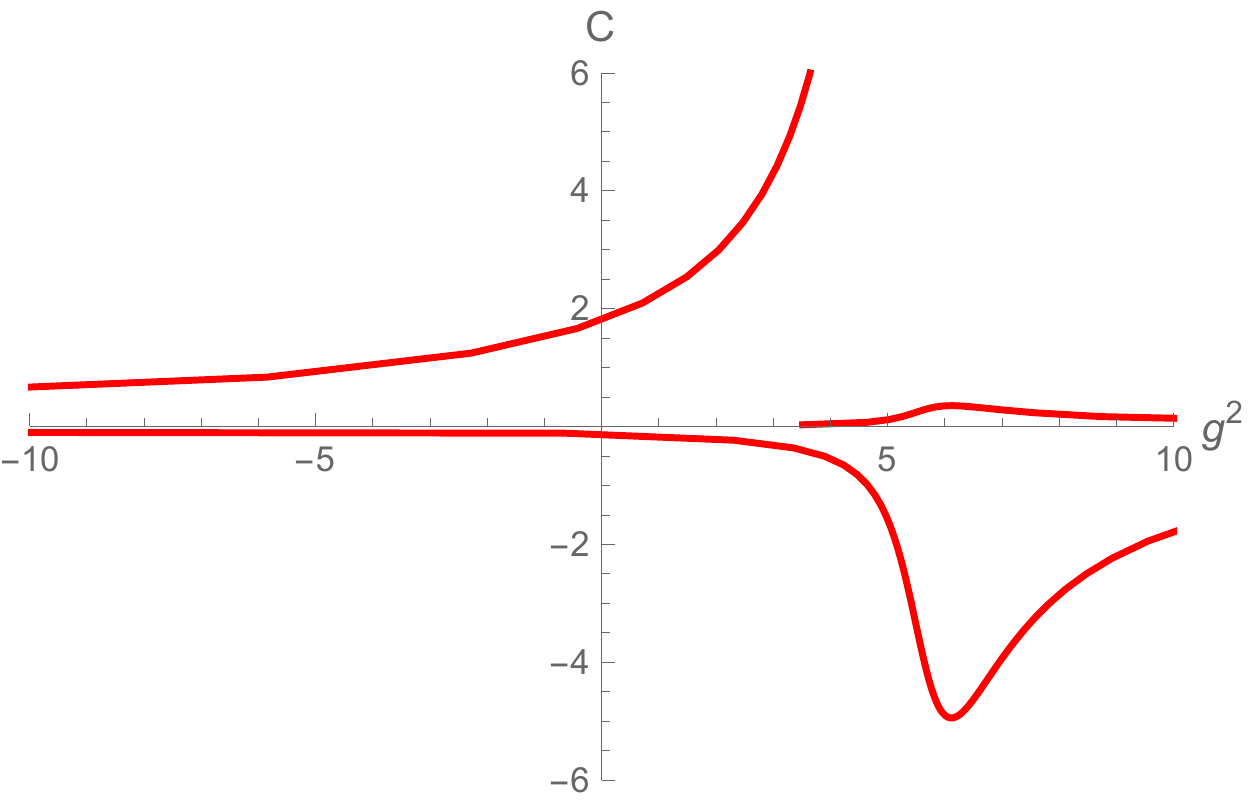}
\caption{Plot of $c_{UV}$ against $g^2$ for the $N_f=9$ theory. }
\label{spiral}
\end{figure}

We can now consider what happens when we switch on an NJL term in this theory.  The easiest method is to use the boundary condition in (\ref{cond}) and convert the $m_{UV}-c_{UV}$ spiral to results in the $c_{UV}-g^2$ plane which we have done for $N_f=9$ in Figure 5. The upper branch of the plot is the most important since these are the stable vacua. Positive $g^2$ enhances $c_{UV}$ from the bare gauge theory values at $g^2=0$.

To understand the role of the other ``arms'' of the spiral it is helpful to plot the effective potential against $m_{UV}$ and $c_{UV}$ which we do in Figure 6. We use the same shading as in the  $m_{UV}-c_{UV}$ spiral for $N_f=9$ in Figure 3. We plot the effective potential against $m_{UV}$ and $c_{UV}$ in the absence of an NJL interaction. Now adding an attractive NJL term corresponds to adding a positive term to the potential $m_{UV}^2 \Lambda^2/g^2$ which we show in the lower plot of Figure 6. The solid curves show the potential with an attractive NJL coupling and  the extra term produces a minimum that grows to larger $m_{UV} (c_{UV})$ as $g^2$ grows.

\begin{figure}[]
\centering
\includegraphics[width=8cm]{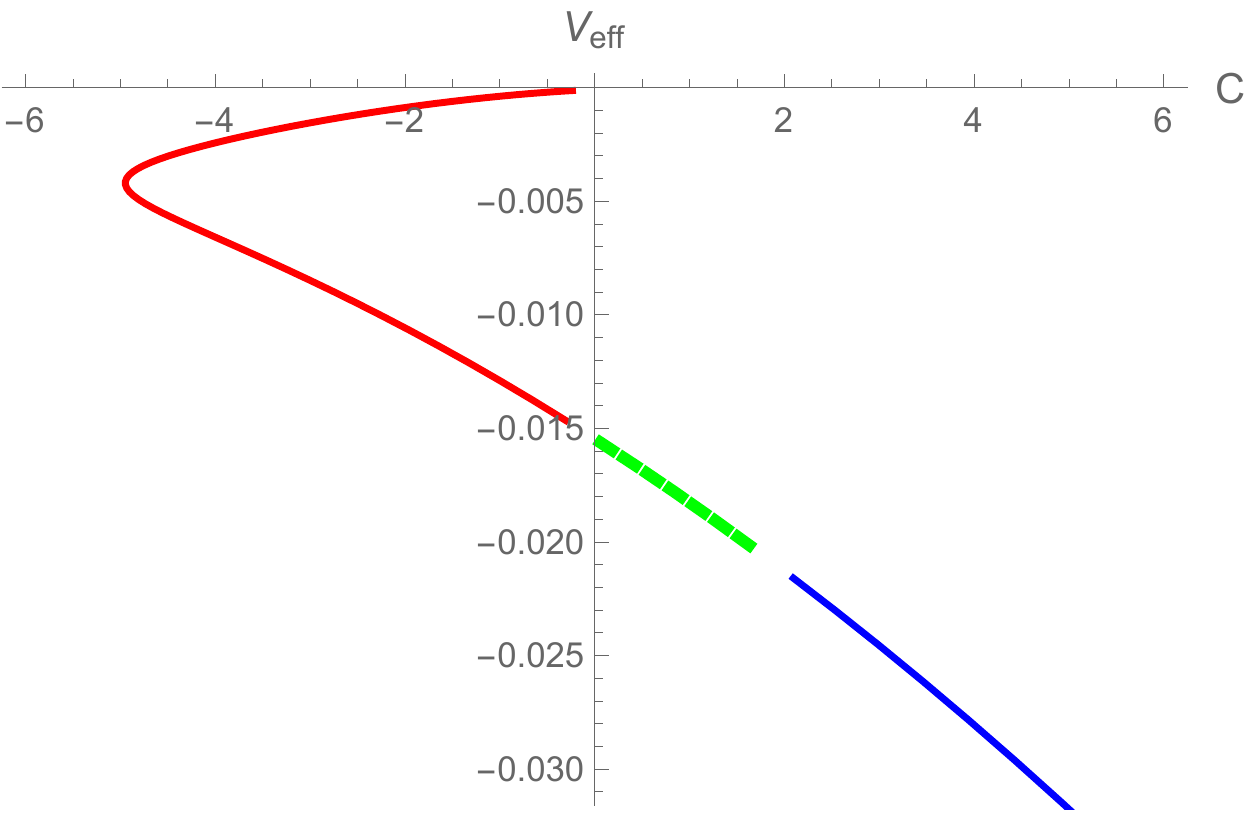}  \includegraphics[width=8cm]{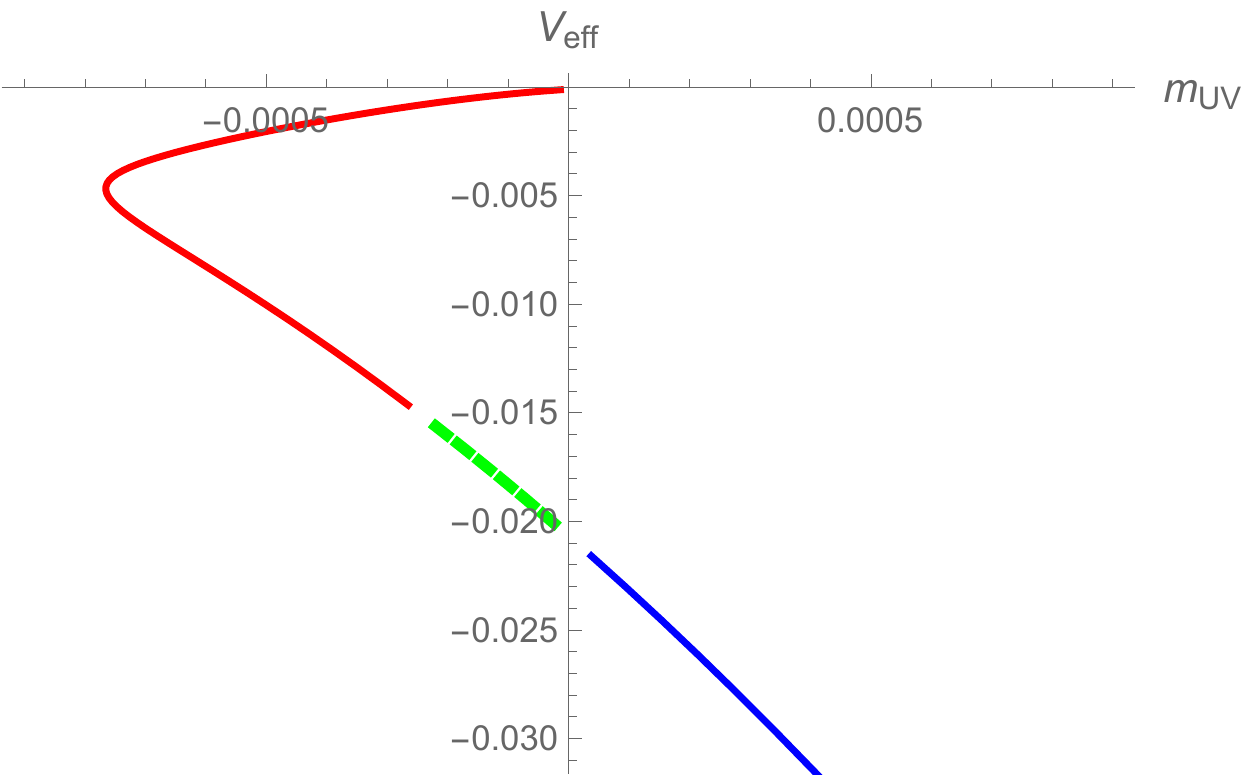} \includegraphics[width=8cm]{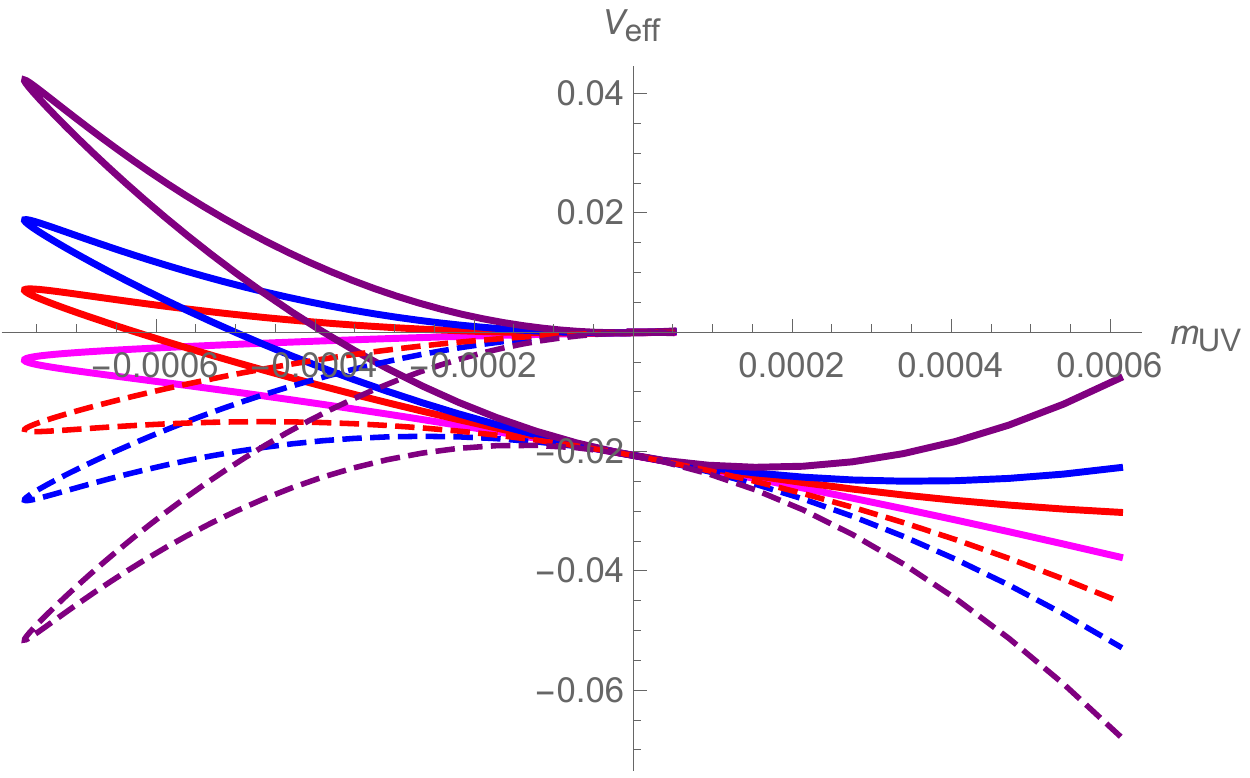}
\caption{The top two plots show the effective action (-S evaluated on the vacuum solutions) for the solutions from Fig 3 for the $N_f=9$ theory as a function of each of $m_{UV}$ and $c_{UV}$.  The lower plot shows the same potential against $m_{UV}$ but including the NJL interaction term. The solid lines are for attractive NJL interactions, dashed lines for repulsive NJL interactions. }
\label{spiral}
\end{figure}

The case of repulsive NJL interactions is also interesting. The $c_{UV}-g^2$ plot in Figure 5 shows that a negative $g^2$ decreases the size of the quark condensate as one might expect, although surprisingly the condensation only fully switches off at infinite repulsive $g^2$. There is a more involved story if we look at these solutions at the level of the  effective potential - see the dashed curves in the lower plot of Figure 6.  As one begins to add a $-m^2 \Lambda^2/g^2$ term to the potential at small $g^2$ the potential is highly unbounded with a single local maximum at $m=0$. The theory has an instability which could presumably be cured by higher dimension operators near the cut off. If we ignore that instability (since an infintessimal repulsion at a very high scale presumably shouldn't totally change the theory), the local maximum matches to the $g^2=0$ vacua of the gauge theory and this presumably represents the impact of the NJL repulsion on the physics of that vacua. As $g^2$ grows the maximum tracks along the curve following the behaviour in the $c_{UV}-g^2$ plot until we arrive at $c_{UV}=0$ but note here that $m_{UV} \neq 0$ and so $g^2$ has diverged. This is related to the existence of the spiral - if the $c_{UV}-m_{UV}$ curve in Figure 3 at negative $m_{UV}$ simply returned to $m_{UV}=c_{UV}=0$ directly then a finite $g^2$ would switch off the condensate. The conclusion is that the more complicated dynamics of the gauge theory (which one could think of as a tower of higher dimension operators) is sufficient to oppose even a very strong repulsive four fermion interaction.

The remaining lower structure in Fig 5 reflects the effect of the NJL term on the unstable vacuum states of the theory. One is tempted to see in Figure 6  metastable vacua but in all cases these states, with higher states of the $\sigma$ condensed, are unstable (the true potential is higher dimensional in the spirit of Fig 4)- they have a tachyonic $\sigma$ or $\pi$ and are smoothly connected to the fully unstable vacuum at infinite positive $m$.

As one increases $N_f$, keeping $\gamma$ fixed at some UV scale across all theories, the spiral rapidly contracts into the origin of the $m_{UV}-c_{UV}$ plane as can be seen from the rapidly sinking axes range in the plots of Figure 3 - this represents the expected reduction in the scale of chiral symmetry breaking as one approaches the BKT transition \cite{Jarvinen:2011qe,Alho:2013dka} at the edge of the conformal window.

\section{$N_f - g^2$ Phase Diagram}

Let us next study the phase structure of the $N_c=3$ theory with $N_f$. Here as we vary $N_f$ we set $\gamma(e) = 2/3 \gamma_*$ so that the step scale $\Lambda_1$ is somewhat comparable in each theory and use a very high UV cut off $\Lambda = e^{20}$ so that all theories have $\gamma \simeq 0$ there. For $N_f < 12$ there is always chiral symmetry breaking triggered by the gauge theory alone. For larger $N_f$, in each case we look at embeddings $L(\rho)$ that correspond to small IR masses (remember we call the IR value of $L$ as $m_{IR}$) .
We extract in the UV $m_{UV}$ and $c_{UV}$ from (\ref{UVform}) and then find the NJL coupling by using (\ref{cond}). Varying the IR mass allows us to plot $m_{IR}$ vs $g$ and extrapolate to the point of the second order transition to find $g_c$ where $m_{IR}$ falls to zero . In this way we can plot the transition in the $g^2-N_f$ plane between theories with conformal IRs and and those with a finite mass gap. We show this phase boundary in Figure 7.

The form of the phase diagram is straightforward - below $N_f=12$ the gauge theory running (in the approximations we use) violates the BF bound and chiral symmetry breaking results.  Above $N_f=12$, the gauge theory alone lies in the conformal window and is ungapped. An NJL term is needed to generate chiral symmetry breaking.  Naively one might think the NJL critical coupling would grow from zero as one moves upwards from $N_f=12$ but in fact our numerical investigations suggest there is a finite critical coupling immediately above the transition leading to a discontinuity in the  plane. It is possible there is an exponential approach to the BKT point at $g^2=0$ but we have not found numerical evidence for it. Note this phase diagram can be compared to the sketch of the similar physics described in \cite{Jarvinen2} where the NJL like behaviour we have found when $N_f > 4 N_c$ is not shown.

\begin{figure}[]
\centering
\includegraphics[width=8cm]{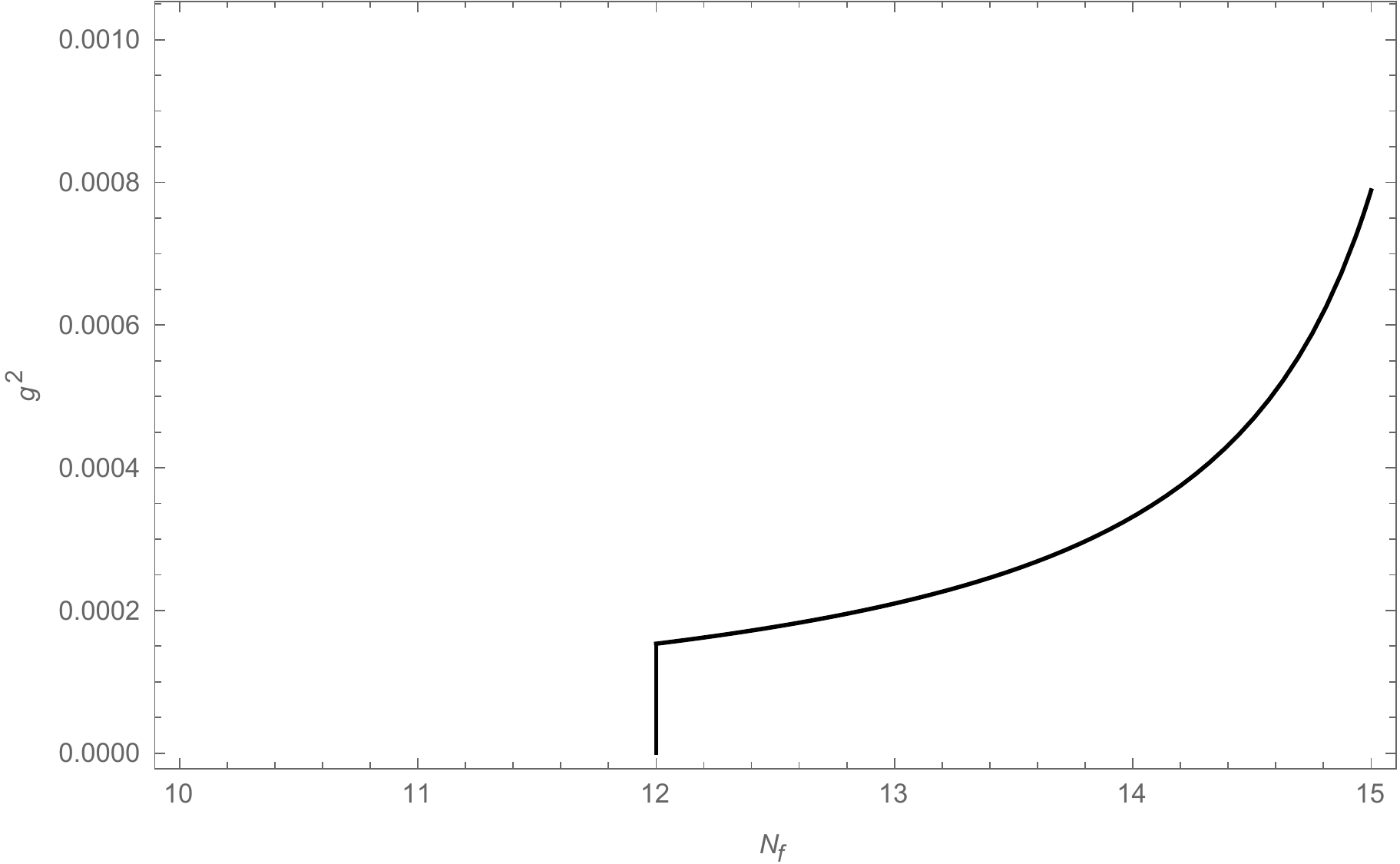}
\caption{The $N_f - g^2$ Phase Diagram - the left hand region has chiral symmetry breaking, the right hand has restored chiral symmetry. }
\label{phase}
\end{figure}

\section{$N_f > 4 N_c$ and Ideal Walking Behaviour}

In the range  $4 N_c < N_f < 11 N_c/2$ the pure gauge theory lives in the conformal window with an IR interacting fixed point but no chiral symmetry breaking. Here we can trigger chiral symmetry breaking by an NJL interaction term.  If the interaction term is tuned so that the mass gap occurs in the IR where there is a large value for $\gamma_*$ then in the UV the condensate is expected to be enhanced by ``walking'' like dynamics as explained in the introduction. We will explore these models in this section.

\subsection{A Simplified Analysis}

Let us first understand how the holographic model rather simply encodes the idea of Ideal Walking. Consider an idealized model with two regimes divided by a sharp scale $\Lambda_1$. At low energies below $\Lambda_1$ the theory has an anomalous dimension $\gamma$ whilst above $\Lambda_1$ $\gamma=0$. The Dynamic AdS/QCD description has $\Delta m^2=0$ in the high energy regime and the constant value $\Delta m^2 = \gamma (\gamma -2)$ in the low energy regime. Now assume a UV NJL interaction triggers chiral symmetry breaking at an energy scale $m_{IR}$ (this in the full model would be the choice of IR boundary condition on the field $L$). If $m_{IR}$ lies above $\Lambda_1$ then the solution for the holographic field $L$ is

\beq L = m + {c \over \rho^2}, \hspace{0.5cm} m \sim m_{IR}, ~~~ c \sim m_{IR}^3 \eeq
where $m,c$ are just fixed by dimensional grounds in terms of the only scale $m_{IR}$ - this a normal ``natural"' theory.

Now imagine moving $m_{IR}$ into the IR regime below $\Lambda_1$. Here the solution looks like
\beq L_{IR} = {\hat{m} \over \rho^\gamma} + {\hat{c} \over \rho^{2- \gamma}}, \hspace{0.5cm} \hat{m} \sim m_{IR}^{1 + \gamma}, ~~~ \hat{c} \sim m_{IR}^{3-\gamma} \eeq
with dimensional analysis again being used to fix the parameters. Now one should evolve this solution to $\Lambda_1$ and match to the UV form of the solutions. In the UV we will have
\beq L_{UV} = m_{UV} + {c_{UV} \over \rho^2}, \hspace{0.5cm} m_{UV} \sim {m_{IR}^{1+\gamma} \over \Lambda_1^\gamma}, ~~~ c_{UV} \sim m_{IR}^{3-\gamma} \Lambda_1^\gamma  \eeq
IR quantities such as $f_\pi$ will be determined simply by dimensional analysis in terms of the IR scale $m_{IR}$ and the UV condensate is relatively enhanced by the presence of $\Lambda_1$ (whilst the UV mass is suppressed).

This is the origin of the effect in the full model we will discuss. A more complete setting is needed to set the UV and IR boundary conditions on the solution and ensure the effective potential from the bulk flows allows the NJL mechanism to operate.  However, if one naively computes the NJL coupling in this approximation one finds $g^2/\Lambda_{UV}^2 = m_{UV}/c_{UV} = m_{IR}^{-2} (m_{IR}/\Lambda)^{2 \gamma}$  which is a constant at $\gamma=1$ and then for fixed $m_{IR}$ rises as $\gamma$ falls. This gives some support to the form of the phase diagram in Fig 7.

\subsection{Two Loop Runnings}

We can now numerically study the more complete theory with the two loop runnings for the $4 N_c < N_f < 11 N_c/2$ theories.  For $N_c=3$ the conformal window lives in the range $12 \leq N_f \leq 15.5$ which corresponds to the fixed point value of $\gamma_{*}$ changing from  1 to 0 as one increases $N_f$. In a previous paper some of the authors \cite{Evans:next} studied the hyper-scaling relations in the holographic model in the absence of NJL terms. Essentially that paper confirmed that the form of the solutions and naive dimensional analysis used in the previous section apply at the level of a percent or better along the flows (because the flows are rather slow and locally taking $\gamma$ to be a constant is a good approximation).

Here we provide a further piece of evidence of the scaling behaviour we expect. Consider the $N_f=13$ theory for  which the running of $\gamma$ is plotted in Figure 1. We fix a UV cut off at the scale where $\gamma(\Lambda_{UV})=0.05$ and then choose a variety of IR initial condition values of $L = m_{IR}$. Solving for $L(\rho)$ we can then extract $m_{UV}$ at the cut off scale from the value of $L(\Lambda_{UV})$. In Figure 8 we plot Log( $m_{UV}/ m_{IR}$) against Log $L_0/\Lambda_{UV}$. If the scaling were the canonical UV scaling then $m_{UV} \simeq m_{IR}$ and the line would be flat at zero. However, we see that as $m_{IR}$ is reduced $m_{UV}$ decreases relative to the canonical scaling expectation and eventually after moving through the running regime of Fig 1 enters a regime where $m_{UV}$ is decreasing with a fixed power as the naive analysis above predicts.

In principle one could perform the same analysis for the condensate but since it is the sub-leading term in the behaviour of $L$ it quite hard to precisely numerically follow it over decades of evolution so we have not produced such clean figures. The naive analysis of the previous section though is clearly appropriate, confirmed for $m$, and the expected growth in the condensate is certainly described in the model.

\begin{figure}[]
\centering
\includegraphics[width=8cm]{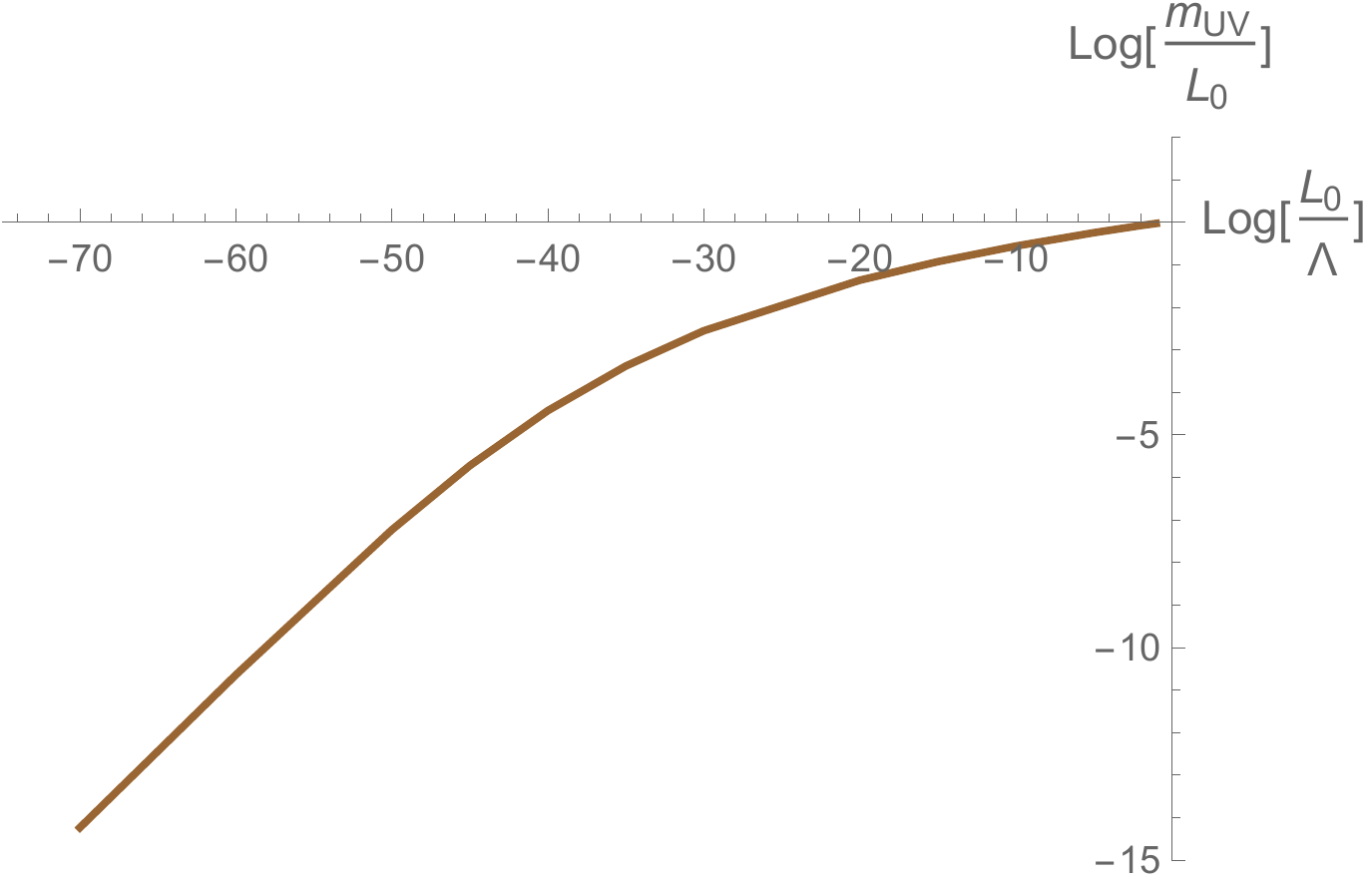}
\caption{Plot of Log $m_{UV}/ m_{IR}$ against Log $m_{IR}/\lambda_{UV}$ for $N_f=13$, $\gamma(\Lambda_{UV})=0.05$}
\label{phase}
\end{figure}
\begin{figure}[]
\centering
\includegraphics[width=8cm]{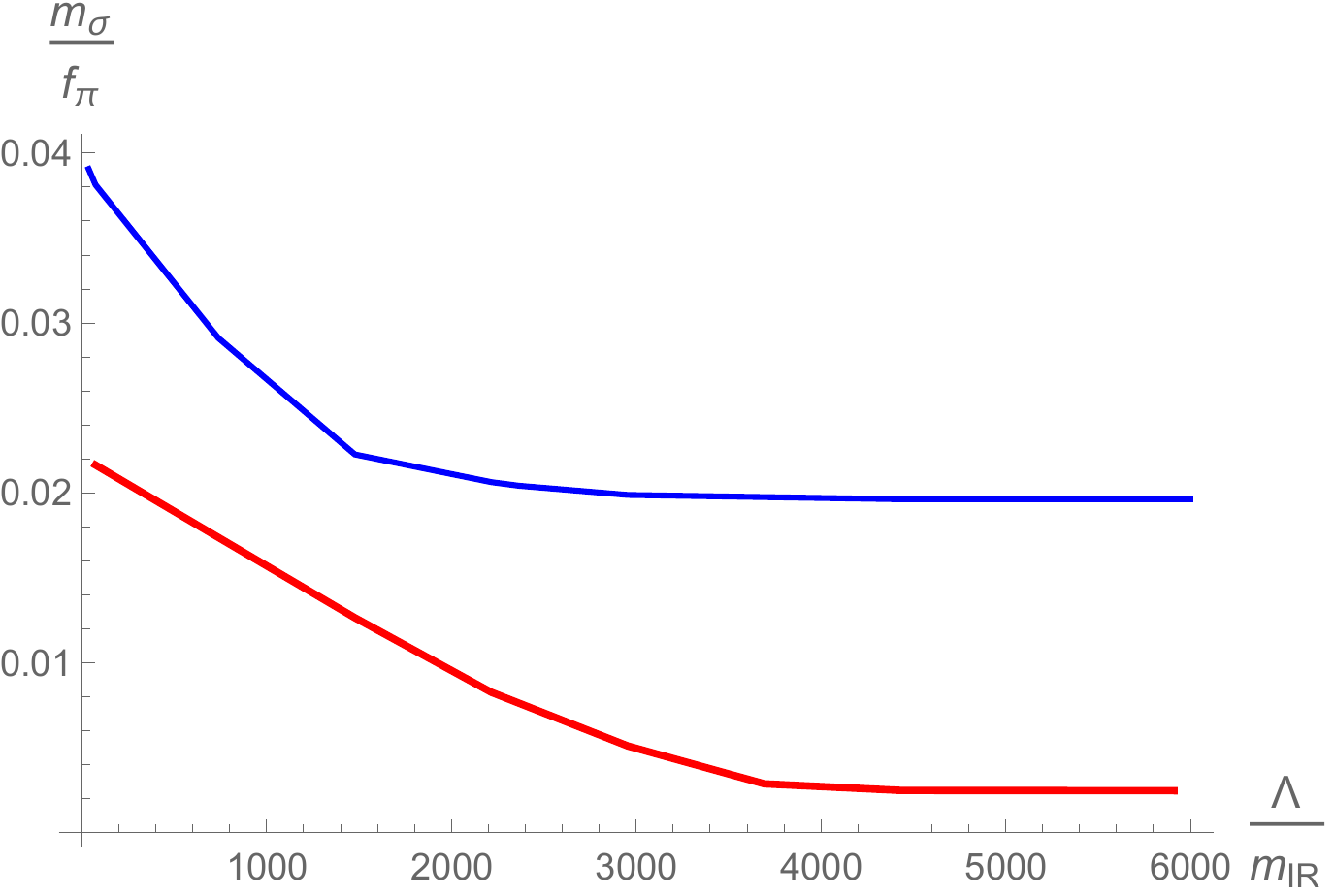}
\caption{The $N_f=13$ theory with $m_{IR}$ lying in the fixed point regime. The $\sigma$'s mass is plotted against Log $\Lambda/ m_{IR}$ for different separations between the IR and UV cut offs.}
\label{phase}
\end{figure}

\subsection{A Light $\sigma$}

The Ideal Walking systems become an interesting possibility for replacements for a normal technicolour description of electroweak symmetry breaking since they enhance the UV condensate which would help to push flavour physics to high scales. A key question though is whether they can describe a light higgs like state (one would need $m_\sigma \simeq f_\pi/2$).

Here again we can provide an analytic answer before we proceed to numerics. The $\sigma$ meson spectrum is found by solving (\ref{scalareom}) although we must be careful with boundary conditions in the prescence of an NJL term. When we find the vacuum form of $L(\rho)$ we interpret the UV boundary constants of the solution as $m_{UV}/c_{UV} = g^2 / \Lambda^2$. When we vary by the field $S$ about that background we must maintain the same value of $g^2 / \Lambda^2$ rather than the usual $S \rightarrow 0$ UV boundary behaviour of the NJL free theory. Numerically this is straightforward.

Now consider (\ref{scalareom}) in the near conformal limit where $\Delta m^2$ varies only very slowly - we may neglect the third term in the equation. Now is there an $M^2=0$ solution? We set the final term to zero also. (\ref{scalareom}) is now precisely (\ref{embedeqn}) which we solved for the background $L_0$. We  already know a  solution, the background $L_0$ itself, that satisfies the relevant NJL boundary condition. So such a massless state is present. This argument shows that if we place the IR mass scale and the UV cut off, separated, but both deep in the IR fixed point regime of the gauge theory we would expect to get an arbitrarily small $\sigma$ mass.

\begin{figure}[]
\centering
\includegraphics[width=8cm]{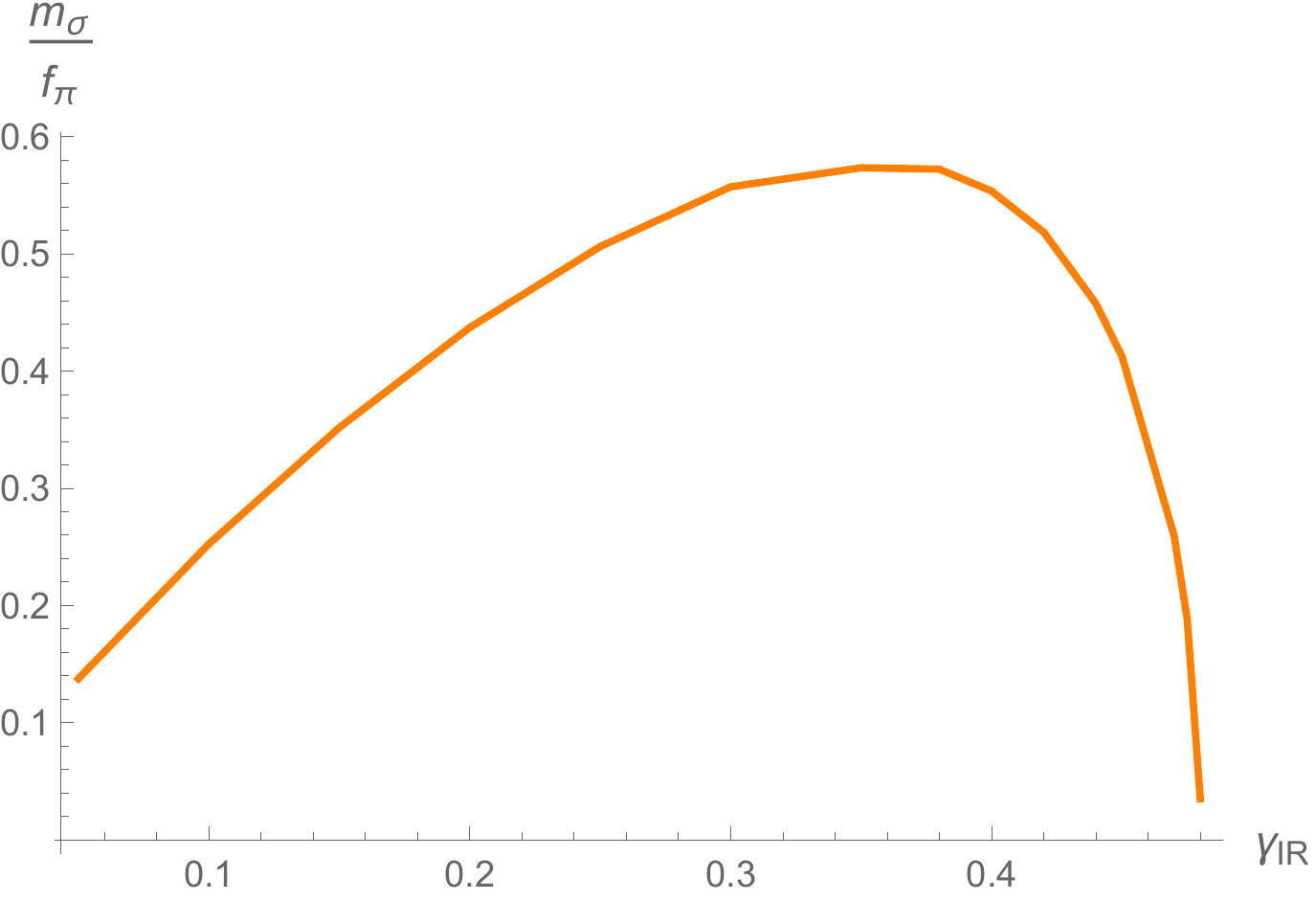}
\caption{This is a plot in the $N_f=12$ theory where the IR fixed point is $\gamma_{IR}=0.48$. Here we have a separation of 7.5 between the $m_{IR}$ and $\Lambda$. We vary $m_{IR}$ to scales with different values of $\gamma_{IR}$ and compute the $\sigma$ mass in units of $f_\pi$. }
\label{phase}
\end{figure}

The set up with both the dynamical scale and the cut off in the deep IR regime generates too small a $\sigma$ mass for an electroweak theory higgs and would also not generate an enhanced condensate because the running does not see a transition in $\gamma$. In fact the runnings in the conformal window are rather slow generically, as can be seen in Figure 1 so $\Delta m^2$ is generically quite flat and the challenge is to make the $\sigma$ as heavy as $f_\pi/2$.

We show some numerical results with the two loop runnings in Figures 9 and 10. In Figure 9 for $N_f=13$ we have fixed the IR mass scale ($m_{IR}$) at a scale in the IR fixed point regime and then varied $\Lambda$ to compute the $\sigma$ mass. The result is small because the coupling is running so slowly.

In Figure 10 we show an example of a theory that achieves a large enough $m_\sigma$  for an electroweak model. Here $N_f=12$ where the IR fixed point is $\gamma_{IR}=0.48$. We have a separation of 7.5 between the scales $m_{IR}$ and $\Lambda$. We vary $m_{IR}$ to scales with different values of $\gamma_{IR}$ and compute the $\sigma$ mass in units of $f_\pi$.   To achieve a larger $m_\sigma$ one needs to sandwich the strongest running between the IR and UV scales.

We conclude that Ideal Walking could, dependent on the precise running at intermediate strength couplings beyond perturbation theory, generate a light $\sigma$ as well as playing the role of enhancing the quark condensate.  In this sense it looks an attractive set up although it relies on NJL terms whose origin is unspecified.

\section{Summary}

We have used a holographic model to study the gauged NJL model with different runnings for the gauge theory in or near the conformal window. We have used the two loop computations of the running of the gauge coupling at $N_c=3$ and varying $N_f$ to represent these runnings from asymptotic freedom to different IR fixed points.

For theories in which $N_f$ lies below $4 N_c$ the runnings for the anomalous dimension of the quark bilinear pass through  $\gamma=1$ and chiral symmetry is triggered when the NJL coupling $g^2$ is zero. Adding an attractive NJL interaction reinforces condensation leading to a bigger mass gap. The basic gauge theories display a spiral pattern in the mass vs condensate plane - at zero quark mass there are vacuum states in which the $\sigma, \sigma^*..$ etc condense although we show only the one with the $\sigma$ alone condensed is stable. This structure is now clearly a prediction of holographic models with symmetry breaking because it has been seen in many models \cite{Filev:2007gb,Jarvinen:2011qe,Jarvinen2,Iqbal:2011aj,BitaghsirFadafan:2018iqr}. Here this structure in the mass-condensate plane impacts when a repulsive NJL term is added, with the surprising result that condensation is only switched off by an infinite NJL coupling.

For $4 N_c < N_f < 11 N_c/2$ the pure gauge theory lies in an IR conformal regime with non-zero $\gamma$. An additional attractive NJL term generates chiral symmetry breaking above a critical NJl coupling value - we have displayed the phase structure. These theories have an intermediate running regime between the $\gamma=0$ UV and the IR fixed point. The values of the UV quark mass and condensate are decreased and increased respectively as the theory runs through this regime. This is the mechanism of Ideal Walking models in which $\langle \bar{q} q \rangle/ f_\pi^3$ can be very much enhanced relative to that in theories with fast running and no IR fixed point. We have also shown that our model predicts a light $\sigma$ particle when the running in these theories is slow which might be helpful in constructing a dynamical model of the electroweak scale.

\vspace{1cm}
\noindent {\bf Acknowledgements:}
NE's work was supported by the
STFC  consolidated  grant  ST/P000711/1  and  WC's  by
an STFC studentship.   KBF  thanks  B.  Robinson
for useful discussions and acknowledges the University of
Southampton for their hospitality during his sabbatical.

\end{document}